# The weakly nonlinear response and non-affine interpretation of the Johnson-Segalman/Gordon-Schowalter model

*Nabil Ramlawi[a], N. Ashwin Bharadwaj[a,1], and Randy H. Ewoldt[a,b,c,*]*

[a] *Department of Mechanical Science and Engineering, University of Illinois at Urbana-Champaign, Urbana, Illinois 61801, USA*

[b] *Materials Research Laboratory, University of Illinois at Urbana-Champaign, Urbana, Illinois 61801, USA*

[c] *Beckman Institute of Advanced Science and Technology, University of Illinois at Urbana-Champaign, Urbana, Illinois 61801, USA*

[1] *Current contact address: Nike, Inc., Beaverton, Oregon 97005, USA*

[*] *Corresponding author:* ewoldt@illinois.edu

(Dated: July 15, 2020)

## Abstract

We derive new analytical solutions for the non-affine Johnson-Segalman/Gordon-Schowalter (JS/GS) constitutive equation with a general relaxation kernel in medium-amplitude oscillatory shear (MAOS) deformation. The results show time-strain separable (TSS) nonlinearity, therefore providing new physically-meaningful interpretation to the heuristic TSS nonlinear parameter in MAOS (Martinetti & Ewoldt *Phys. Fl.* (2019)). The upper-convected, lower-convected, and corotational Maxwell models are all subsets of the results presented here. The model assumes that the microscale elements causing stress in the material slip compared to the continuum deformation. We introduce a visualization of the non-affine deformation field that acts on stress-generating elements to reinforce the physical interpretation of the JS/GS class of models. Finally, a case study is presented where previously published results, from fitting TSS models to MAOS data, can be re-interpreted based on the concept of non-affine motion of the JS/GS framework.

**Keywords**: Medium-amplitude oscillatory shear, MAOS, Large-amplitude oscillatory shear, LAOS, Non-affine deformation, Polymers, Nonlinear rheology, Constitutive models

## I. INTRODUCTION

Weakly-nonlinear analysis of complex fluid rheology is excellent for relating rheology to structure and studying material-level physics of fluids. Weakly-nonlinear rheological characterization, such as medium-amplitude oscillatory shear (MAOS) (Paul 1969; Onogi et al. 1970; Davis and Macosko 1978; Hyun and Wilhelm 2009; Wagner et al. 2011; Ewoldt and Bharadwaj 2013; Bharadwaj and Ewoldt 2014, 2015a, b) and medium-amplitude parallel superposition MAPS (Lennon et al. 2020), produces more information than linear viscoelastic measures while staying mathematically tractable to theoretical prediction. For example, recently our group collaborated (Martinetti et al. 2018) to settle a 70-year debate and infer the nonlinear mechanisms of an aqueous viscoelastic liquid reversible polymer network by combining MAOS measurements with a novel asymptotically nonlinear viscoelastic model (Bharadwaj et al. 2017) and the Polymer Reference Interaction Site Model (PRISM) (Schweizer and Curro 1994).

Here we derive new analytical predictions for MAOS for a nonlinear viscoelastic fluid constitutive model, wherein the nonlinear parameter (which can be fit to experimental observations) can be interpreted via non-affine deformation of the material structures that "slip" compared to the continuum deformation. Specifically, we consider the Johnson-Segalman (JS) integral model (Johnson and Segalman 1977), which is equivalent, in the single mode Maxwell relaxation limit, to the Gordon-Schowalter (GS) derivative applied to a Maxwell model (Gordon and Schowalter 1972; Larson 1988). Several MAOS solutions for various models are recovered as subsets of this JS/GS framework by tuning the nonlinear parameter (upper-convected, lower-convected, and corotational Maxwell (Giacomin et al. 2011)).

The results derived here provide a new model to be considered when fitting MAOS data, adding to the existing toolbox of known analytical results (Bharadwaj and Ewoldt 2015a; Saengow et al. 2017). Not only does this support better checks on model credibility when fitting by enabling multiple possible models to fit rheological data (Freund and Ewoldt 2015), but it provides a physical interpretation for inference of the fluid physics, in contrast to otherwise heuristic existing models including the corotational Maxwell model (Giacomin et al. 2011) and generalized time-strain separable (TSS) integral models (Martinetti and Ewoldt 2019). Recently, (Song et al. 2020) expanded the library of MAOS solutions extensively by analytically deriving the MAOS material functions for several nonlinear viscoelastic constitutive models, including the JS/GS model in the special case of single mode relaxation. Here, we derive the analytical results of the general JS integral model that allows for any type of relaxation function and hence offers more flexibility in fitting data and expands the list of materials that can be studied using this model. For example, a power law relaxation function $G(t) = St^{-n}$ that is observed for critical gels (Chambon and Winter 1987), can only be modeled with JS/GS non-affine deformation using results presented here. In addition, our work here offers a detailed analysis that illustrates the physical interpretation of this model and its non-affine deformation.

The outline of our contributions is as follows. We theoretically derive the MAOS intrinsic material functions for the JS/GS model and show the results for the single mode relaxation case. In the discussion, we first show that the model is time-strain separable (TSS), thus providing a route to interpret, at least for a certain range of the nonlinear TSS parameter, the general TSS class of models that was not found before. We continue the discussion by introducing a visualization of the non-affine deformation field which acts on stress-generating elements to reinforce the physical interpretation of the JS/GS class of models. The discussion also includes a comment on interpreting the model physically and how it applies to different material classes. Finally, a case study is presented where previously published results, from heuristically fitting TSS models to MAOS data, can be re-interpreted based on the concept of non-affine motion of the JS/GS framework.

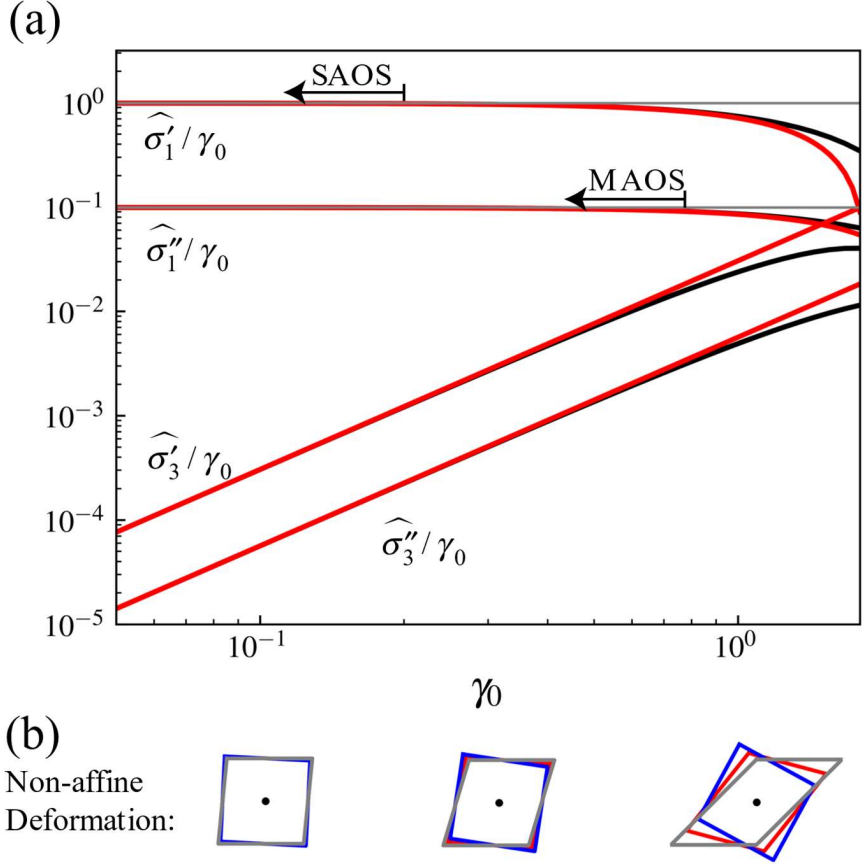

Figure 1: Large-amplitude oscillatory shear (LAOS) response of the JS/GS model. (a) Strain amplitude sweep at De = 10 with nonlinear slip parameter $a = 0.5$. Stress harmonics for the full solution (black) obtained numerically and the truncated MAOS expansion (red) derived analytically in this work (Section III). Limits of SAOS and MAOS ranges are shown at the first occurrence of 1% deviation between each truncation and the fully nonlinear solution. The normalized harmonics are defined as $\widehat{\sigma}'_n = \sigma'_n/G_0$ and $\widehat{\sigma}''_n = \sigma''_n/G_0$. (b) Visualization of the fully nonlinear non-affine deformation of the stress-generating material elements compared to the affine (gray) for De = 10 and $\gamma_0 = \{0.1, 0.3, 1\}$, based on results in Section I.B. (blue) $a \to 0$ equivalent to corotational Maxwell, (red) $a = 0.5$, and (gray) affine equivalent to $a = 1$ i.e. upper-convected.

## II. BACKGROUND

A.   Medium-amplitude oscillatory shear (MAOS)

Oscillatory deformation is preferred for weakly-nonlinear rheometry, compared to step input forcing, since oscillations lock in at each frequency (Deborah number) to provide a high signal-to-noise ratio for the weak nonlinear signals. Imposing a simple shear strain input $\gamma(t) = \gamma_0 \sin \omega t$, the shear stress response $\sigma_{21}$ can be written as (Ewoldt 2013)

$$\sigma_{21}(t;\omega,\gamma_0) = \sum_{n}^{\infty} \{\sigma'_n(\omega,\gamma_0)\sin n\omega t + \sigma''_n(\omega,\gamma_0)\cos n\omega t\}, \qquad (1)$$

where $\sigma'_n$ and $\sigma''_n$ are Fourier coefficients. Even harmonics of $n$ are excluded for time-periodic shear-symmetric responses, for which stress is an odd function of strain and only $n$=odd are required. The linear (small-amplitude oscillatory shear, SAOS) and weakly-nonlinear MAOS regimes at any frequency $\omega$ are defined by a power-series expansion with respect to the amplitude ($\gamma_0$ or $\dot{\gamma}_0 = \gamma_0\omega$ ) (Cho et al. 2005; Ewoldt et al. 2008; Hyun et al. 2011; Ewoldt and Bharadwaj 2013) for example (Ewoldt and Bharadwaj 2013)

$$\begin{aligned}
\sigma'_1(\omega,\gamma_0) &= G'\gamma_0 + G'_{13}\gamma_0^3 + ... = G'\gamma_0 + \phantom{\omega}[e_1]\gamma_0^3 + ...\\
\sigma''_1(\omega,\gamma_0) &= G''\gamma_0 + G''_{13}\gamma_0^3 + ... = G''\gamma_0 + \omega[v_1]\gamma_0^3 + ...\\
\sigma'_3(\omega,\gamma_0) &= \phantom{G'\gamma_0} + G'_{33}\gamma_0^3 + ... = \phantom{G'\gamma_0} - [e_3]\gamma_0^3 + ...\\
\sigma''_3(\omega,\gamma_0) &= \phantom{G'\gamma_0} + G''_{33}\gamma_0^3 + ... = \phantom{G'\gamma_0} + \omega[v_3]\gamma_0^3 + ...
\end{aligned} \qquad (2)$$

where the coefficients of the expansion depend on frequency but not amplitude. Here $G'(\omega)$ and $G''(\omega)$ are the two (2) linear viscoelastic moduli, and four (4) parameters are required to fully describe the asymptotic deviation from linearity: two for the first-harmonic response, and two for the third-harmonic response.[*]

The measured stress harmonics are often plotted as a function of strain amplitude, as shown in Figure 1 for the model we consider in this paper (see Section II.B and III for model details). When normalized by strain amplitude, the first harmonics approach asymptotic plateaus equal to the linear viscoelastic moduli $G'$ and $G''$ in the limit of amplitude $\to 0$. At any non-zero strain amplitude, nonlinearities are always present, in the sense of a power-law expansion. These ever-present nonlinearities cause a deviation from the linear response first-harmonic plateau and additionally generate third-harmonics in the shear stress signal, as in Eq.(2). The labeled SAOS limit in Figure 1 is defined by the domain of strain amplitude where all normalized harmonics of the fully nonlinear solution deviate by no more than 1% from the linear plateau. Similarly, the weakly-nonlinear MAOS regime, which includes only the first deviation from linearity, is defined by comparing the deviation of the fully nonlinear solution to the truncated MAOS expansion (up to order $\gamma_0^3$). The choice of 1% deviation is a subjective choice (0.1% or 10% may also be reasonable) and depends on the accuracy and resolution of the experimental measurement, theoretical model, or numerical simulation.

---

[*] Here we have used an expansion with respect to strain amplitude, $\gamma_0^3$, but this is a subjective choice and expansions can also be made with respect to strain-rate amplitude $\dot{\gamma}_0^3$, e.g. see (Giacomin et al. 2011), their Equation 9. Non-integer power expansions have also been observed experimentally (Natalia et al. 2020) though with limited theoretical prediction (Blackwell and Ewoldt 2016).

The four MAOS measures (expansion coefficients) can be represented in various ways. Two options are shown in Eq.(2), the latter of which involves the Chebyshev expansion coefficients $[e_1](\omega), [v_1](\omega), [e_3](\omega)$ and $[v_3](\omega)$ as defined in (Ewoldt and Bharadwaj 2013), where "e" is for elastic with SI units [Pa], "v" is for viscous with SI units [Pa·s], and a notable negative sign appears in front of $[e_3]$ due to the conversion of Chebyshev to Fourier coefficients (Ewoldt et al. 2008). The Chebyshev expansion coefficients describe oscillatory waveforms visualized as Lissajous-Bowditch curves (hysteresis loops) of stress-versus-strain and stress-versus-rate, but it is convenient to use the Fourier representation in Eq.(2) for signal processing. The Chebyshev coefficients offer physical interpretation for elastic and viscous nonlinearities of all four MAOS nonlinearities (see (Ewoldt and Bharadwaj 2013), their Figures 6-7), e.g. in contrast to time-domain Fourier coefficients or magnitude nonlinearities alone (Hyun and Wilhelm 2009). The MAOS material functions of Eq.(2) can also be related to an asymptotic power expansion of the SPP LAOS framework of Rogers and co-workers, as recently shown in (Choi et al. 2019), their Equations (F1)-(F2), which clearly show that the SPP metrics identify so-called "elastic" and "viscous" effects differently than the Chebyshev expansion of Eq.(2).[†]

---

[†] For example, a *purely elastic* weakly-nonlinear response $\sigma_{21}(\gamma)$ will generate only elastic coefficients $G'$, $[e_1]$, and $[e_3]$ in the Chebyshev expansion, whereas it generates non-zero $G'$, $G_t$, and "viscous" $\eta_t$ in the SPP framework. Similarly, a purely viscous nonlinear response $\sigma_{21}(\dot{\gamma})$ generates only viscous Chebyshev expansion coefficients, but generates $\eta'$ and both $\eta_t$ and "elastic" $G_t$ in SPP.

## B. Constitutive Model

We consider the Johnson-Segalman (JS) non-affine deformation model (Johnson and Segalman 1977), or equivalently the generalized Gordon-Showalter (GS) (Gordon and Schowalter 1972) Maxwell model.

Gordon and Schowalter modified the molecular theory of elastic dumbbells (simplified polymer strands) to allow non-affine deformation by adding slip between the velocity gradient felt by the dumbbells and the gradient imposed by the continuum deformation. The resulting continuum level stress-strain constitutive equation, derived by (Gordon and Everage 1971), is given by

$$\tau_0 \overset{\square}{\underline{\underline{\sigma}}} + \underline{\underline{\sigma}} = \eta_0 \underline{\underline{\dot{\gamma}}} \qquad (\eta_0 \equiv G_0 \tau_0), \qquad (3)$$

where $\overset{\square}{\underline{\underline{\sigma}}}$ is the Gordon-Schowalter convected derivative defined (Larson 1988) as

$$\overset{\square}{\underline{\underline{\sigma}}} = \frac{D}{Dt}\underline{\underline{\sigma}} - \underline{\underline{\omega}}^T \cdot \underline{\underline{\sigma}} - \underline{\underline{\sigma}} \cdot \underline{\underline{\omega}} - \frac{a}{2}\left(\underline{\underline{\dot{\gamma}}} \cdot \underline{\underline{\sigma}} + \underline{\underline{\sigma}} \cdot \underline{\underline{\dot{\gamma}}}\right) \qquad (4)$$

where $D/Dt = \partial/\partial t + \underline{v} \cdot \underline{\nabla}$ is the substantial derivative, $\underline{\underline{\omega}} = \tfrac{1}{2}\left(\underline{\nabla}\underline{v} - (\underline{\nabla}\underline{v})^T\right)$ is the vorticity tensor, and $\underline{\underline{\dot{\gamma}}} = \underline{\nabla}\underline{v} + (\underline{\nabla}\underline{v})^T$ is the rate of deformation tensor in the continuum velocity field $\underline{v}(\underline{x},t)$. This constitutive model is equivalent to the Maxwell model, but with replacing the time derivative with the GS convected derivative, and therefore it is referred to as the GS Maxwell model, which is a generalization of the Upper-convected Maxwell model.

A different but equivalent route used by Johnson and Segalman was to use the Lodge integral equation and replace the velocity gradient with a non-affine velocity gradient, eventually resulting in an integral form model given by

$$\underline{\underline{\sigma}} = \int_{-\infty}^{t} G_A(t-t')\underline{\underline{S}}(a;t,t')dt' \qquad (5)$$

where $\underline{\underline{S}}$ is an alternative rate of strain measure defined in Section III, and $G_A(t)$ is a time-dependent kernel function that is equivalent to the stress relaxation modulus in the affine limit of the model. While Gordon and Schowalter assumed a polymer in solution system, Johnson and Segalman did not restrict their model to a specific microstructure as will be discussed more in Section IV. Moreover, the GS Maxwell model in Eq. (4) is a subset of the JS model in Eq.(5). In this work, we will use this integral form to arrive at the MAOS solutions for its simplicity and generality. There seems to be a naming confusion in the literature (Larson 1988; Radulescu and Olmsted 2000; Song et al. 2020), where the single-mode GS Maxwell model is sometimes misnamed as the JS model. The mathematical response of the single-mode GS differential model is identical to the single-mode JS integral model, but the JS model in integral from is extremely general, allowing any relaxation kernel including those that are not easily represented as a sum of exponentials. We hope the background information presented below and in Section III clarifies the different origins of the two models.

In both cases, GS and JS, the non-affine motion can be interpreted as introducing a non-affine "effective" velocity gradient tensor to be used for the stress calculator that is given by

$$\underline{\underline{L}} = \left(\frac{a+1}{2}\right)(\underline{\nabla v}) + \left(\frac{a-1}{2}\right)(\underline{\nabla v})^T, \tag{6}$$

where $a$ is the non-linear model parameter. In Eq.(6), the convention used to evaluate components of the velocty gradient tensor $\underline{\nabla v}$ is $\nabla_i v_j = \partial v_j / \partial x_i$ (unlike the definition in the work of (Johnson and Segalman 1977)). The deformation resulting from this gradient, illustrated in Figure 1, "slips" compared to the affine velocity field but still involves stretching and rotation (or only rotation in the limit of $a = 0$, equivalent to the corotational Maxwell model). The affinity parameter (or slip parameter) $a$ is the only nonlinear parameter of the model. In various limits, it recovers other known models as a subset (Larson 1988), e.g. $a = 1$ recovers the Lodge integral model or upper-convected Maxwell (UCM) differential model, $a = -1$ is the lower-convected Maxwell model, and $a = 0$ is the corotational Maxwell (CM) model (Larson 1988).

Additional model details are included as needed in Sections III and IV.

## III.    RESULTS: THEORETICAL DERIVATION OF MAOS SIGNATURES

Before deriving the MAOS material functions of the JS/GS model, a brief outline of the procedure is given here.  First, an oscillatory homogeneous simple shear velocity field is imposed as an input to the constitutive model to compute the output shear stress as a function of time. This equation of stress was derived in the work of Johnson & Segalman, and we repeat the steps below to avoid confusion related to change in notation. In particular, here we use the rate of strain tensor $\underline{\underline{\dot{\gamma}}}$ instead of the deformation tensor $D = \frac{1}{2}\underline{\underline{\dot{\gamma}}}$, and we use $G_A$ as the affine relaxation modulus instead of $G = 2G_A$. We prefer this notation because $G_A$ is the actual relaxation modulus in the affine limit, while $G$ in (Johnson and Segalman 1977) is off by a constant. Next, the resultant stress solution is expanded in amplitude and in frequency. The order of the expansions does not change the results, but here we choose to expand first in amplitude (up to third order, i.e. $\gamma_0^3$) for mathematical convenience. To expand in frequency, the Euler-Fourier equations are applied to the amplitude-expanded stress solution to obtain the stress harmonics defined in Eq. (1) up to third order in $\gamma_0$. Using the result, the MAOS material functions are readily identified. The choice of MAOS material functions to be reported is not unique, but rather depends on the chosen representation of the stress expansion in both frequency and amplitude. Here we adopt the MAOS material functions defined by Ewoldt and Bharadwaj, which are related to other representations in literature (Ewoldt and Bharadwaj 2013).

### A.  General relaxation kernel result

Following the above procedure, we start by assuming a one-dimensional homogeneous simple shear flow and represent the components of the Cartesian velocity field as

$$\underline{v} = \begin{bmatrix} v_x \\ v_y \\ v_z \end{bmatrix} = \begin{bmatrix} \dot{\gamma}(t)y \\ 0 \\ 0 \end{bmatrix} \tag{7}$$

where $\dot{\gamma}(t)$ is the rate of strain. The rate of strain tensor is calculated as $\underline{\underline{\dot{\gamma}}} = \underline{\nabla v} + (\underline{\nabla v})^{\mathrm{T}}$, furnishing

$$\underline{\underline{\dot{\gamma}}} = \dot{\gamma}(t) \begin{pmatrix} 0 & 1 & 0 \\ 1 & 0 & 0 \\ 0 & 0 & 0 \end{pmatrix}. \tag{8}$$

The choice of $\gamma(t) = \gamma_0 \sin \omega t$ follows the convention of an oscillatory characterization protocol (Ewoldt 2013), giving the strain rate input of the form

$$\dot{\gamma}(t) = \gamma_0 \omega \cos \omega t. \tag{9}$$

To compute the stress we use the integral form of Johnson and Segalman from Eq.(5). The alternative deformation measure is defined as

$$\underline{\underline{S}}(t,t') \equiv \underline{\underline{E}}(t,t') \cdot \left(a\underline{\underline{\dot{\gamma}}}(t')\right) \cdot \left(\underline{\underline{E}}(t,t')\right)^T, \tag{10}$$

where $\underline{\underline{E}}(t,t')$ is obtained by solving the differential equation

$$\frac{D}{Dt'} \underline{\underline{E}}(t',t) = -\underline{\underline{E}}(t',t) \cdot \left(\underline{\underline{L}}\right)^T \tag{11}$$

with the initial condition $\underline{\underline{E}}(t,t') = \underline{\underline{I}}$. It is important to notice that in the linear limit of small strains, Eq. (5) reduces to the stress equation given by the Boltzmann superposition as

$$\underline{\underline{\sigma}} = \int_{-\infty}^{t} G_A(t-t') \left(a\underline{\underline{\dot{\gamma}}}(t')\right) dt' \tag{12}$$

with the effective "Non-affine" (NA) strain-rate being $\underline{\underline{\dot{\gamma}}}_{NA}(t') = a\underline{\underline{\dot{\gamma}}}(t')$. Moreover, the stress equation reduces to the Lodge equation the affine limit of $a = 1$ (Larson 1988).

For homogeneous unsteady simple shear flow (as defined in Eq.(7)), the set of differential equations in Eq.(11) can be solved to find the components of $\underline{\underline{E}}(t,t')$ (see (Johnson and Segalman 1977), but note the difference in notation for $\dot{\gamma}(t)$ [‡]), resulting in

$$\underline{\underline{E}}(t,t') = \begin{pmatrix} \cos \lambda s(t,t') & \sqrt{\dfrac{1+a}{1-a}} \sin \lambda s(t,t') & 0 \\ -\sqrt{\dfrac{1-a}{1+a}} \sin \lambda s(t,t') & \cos \lambda s(t,t') & 0 \\ 0 & 0 & 1 \end{pmatrix} \quad (13)$$

where $\lambda = \tfrac{1}{2}\sqrt{1-a^2}$ and $s(t,t') = \int_{t'}^{t} \dot{\gamma}(t'')dt''$. The components of the alternative rate of strain measure are then calculated from Eq.(10) as

$$\underline{\underline{S}}(t,t') = a\dot{\gamma}(t') \begin{pmatrix} \left(\sqrt{\dfrac{1+a}{1-a}}\right)\sin 2\lambda s(t,t') & \cos 2\lambda s(t,t') & 0 \\ \cos 2\lambda s(t,t') & -\left(\sqrt{\dfrac{1-a}{1+a}}\right)\sin 2\lambda s(t,t') & 0 \\ 0 & 0 & 0 \end{pmatrix}. \quad (14)$$

---

[‡] We changed the symbol for shear-rate from $k$ to $\dot{\gamma}$.

Equation (14) holds for any unsteady homogeneous simple shear defined by $\dot{\gamma}(t)$. It is now possible to calculate the shear stress from the 21-component of $\underline{\underline{S}}(t,t')$ from Eq.(14) as

$$\sigma_{21} = \int_{-\infty}^{t} G_A(t-t') S_{21}(t,t') dt' \quad (15)$$

that can be rewritten using Eq.(14) as

$$\sigma_{21}(t) = a \int_{-\infty}^{t} \left\{ G_A(t-t') \dot{\gamma}(t') \cos\left( \sqrt{1-a^2} \int_{t'}^{t} \dot{\gamma}(t'') dt'' \right) \right\} dt'. \quad (16)$$

This same result was obtained in the work of Johnson-Segalman in their Eq. (3.8) (Johnson and Segalman 1977), noting the difference in notation as previously mentioned, in addition to a factor of $\tfrac{1}{2}$ showing up there due to the definition of $G$ they made in their Eq. (2.25).

For an oscillatory deformation field defined in Eq.(9) the integral involving the shear rate in Eq.(16) is evaluated and the following expression is obtained for the shear stress

$$\sigma_{21}(t) = a \int_{-\infty}^{t} \left\{ G_A(t-t')(\gamma_0 \omega \cos \omega t') \cos\left(\gamma_0 \sqrt{1-a^2}\left[\sin \omega t - \sin \omega t'\right]\right) \right\} dt'. \tag{17}$$

Here we assumed the oscillations started at $t = -\infty$ to obtain the steady time-periodic (alternance state) shear stress response. To evaluate the time integral in Eq.(17), it is convenient to carry out a variable transformation $t - t' = s$, that reduces it to

$$\sigma_{21}(t) = a\gamma_0 \omega \left[ \int_0^\infty \left\{ G_A(s) \cos \omega(t-s) \cos\left(\gamma_0 \sqrt{1-a^2}\left[\sin \omega t - \sin \omega(t-s)\right]\right) \right\} ds \right]. \tag{18}$$

At this point, we reach the fully nonlinear time-dependent solution, and the necessary expansions are taken next. In the asymptotic limit of small strain amplitude, $\gamma_0 \to 0$, it is possible to expand the cosine term involving $\gamma_0$ in Eq.(18) as a Taylor series about $\gamma_0 = 0$, resulting in a simplified expression

$$\sigma_{21}(t) = a\gamma_0 \omega \left[ \int_0^\infty \left\{ G_A(s) \cos \omega(t-s) \left[ 1 - \frac{1}{2!}\gamma_0^2(1-a^2)\{\sin \omega t - \sin \omega(t-s)\}^2 + O(\gamma_0^4) \right] \right\} ds \right]. \tag{19}$$

Applying the Euler-Fourier equations results in the expanded stress harmonics which are coefficients of the $\cos \omega t$, $\sin \omega t$, $\cos 3\omega t$, $\sin 3\omega t$ terms. This operation further simplifies Eq.(19) and allows the expression of shear stress as a third order power series expansion in strain amplitude $\gamma_0$, where each order is subsequently separated into orthogonal harmonics as

$$\sigma_{21}(t) = \gamma_0 \left\{ \left[ \omega \int_0^\infty aG_A(s) \sin \omega s \, ds \right] \sin \omega t + \left[ \omega \int_0^\infty aG_A(s) \cos \omega s \, ds \right] \cos \omega t \right\}$$

$$+ \gamma_0^3 \left\{ \begin{array}{l} \left[ \dfrac{1}{4}(1-a^2) \int_0^\infty aG_A(s) \omega [-2\sin \omega s + \sin 2\omega s] ds \right] \sin \omega t \\[6pt] + \left[ \dfrac{1}{4}(1-a^2) \int_0^\infty aG_A(s) [-\cos \omega s + \cos 2\omega s] ds \right] \cos \omega t \\[6pt] + \left[ \dfrac{1}{8}(1-a^2) \int_0^\infty aG_A(s) \omega [\sin \omega s - 2\sin 2\omega s + \sin 3\omega s] ds \right] \sin 3\omega t \\[6pt] + \left[ \dfrac{1}{8}(1-a^2) \int_0^\infty aG_A(s) [\cos \omega s - 2\cos 2\omega s + \cos 3\omega s] ds \right] \cos 3\omega t \end{array} \right\}. \quad (20)$$

$$+ O(\gamma_0^5)$$

The SAOS and MAOS functions are identified by comparing Eq. (20) to the shear stress expansion given in Eq.(1)–(2). The linear viscoelastic material functions, e.g. storage and loss moduli, appear as coefficients with the first power of $\gamma_0$

$$G'(\omega) = \omega \int_0^\infty aG_A(s) \sin \omega s \, ds, \quad (21)$$

$$G''(\omega) = \omega \int_0^\infty aG_A(s) \cos \omega s \, ds. \quad (22)$$

By comparing Eqs. (21)–(22) to the definition of the linear storage and loss moduli, we can deduce that the effective stress relaxation modulus in the non-affine JS/GS model is given by

$$G(s) = aG_A(s). \quad (23)$$

The relaxation modulus $G(s)$ is well-behaved and finite in the limit of $a \to 0$, but this requires the magnitude of the affine relaxation modulus $G_A$ to tend to $\infty$ in this limit that is equivalent to the corotational Maxwell model (Johnson and Segalman 1977). Here we choose to follow the convention of (Ewoldt 2013), and use the intrinsic measures of MAOS as defined in Eq.(2), which for the JS model with generic kernel $G(t)$ are

$$[e_1](\omega) = G'_{13} = \frac{1-a^2}{4} \int_0^\infty G(s)\omega[-2\sin\omega s + \sin 2\omega s]ds$$
$$= \frac{1-a^2}{8}[-4G'(\omega) + G'(2\omega)] \qquad (24)$$

$$[v_1](\omega) = \frac{G''_{13}}{\omega} = \frac{1-a^2}{4} \int_0^\infty G(s)[-\cos\omega s + \cos 2\omega s]ds$$
$$= \frac{1-a^2}{8\omega}[-2G''(\omega) + G''(2\omega)] \qquad (25)$$

$$[e_3](\omega) = -G'_{33} = -\frac{1-a^2}{8} \int_0^\infty G(s)\omega[\sin\omega s - 2\sin 2\omega s + \sin 3\omega s]ds$$
$$= \frac{1-a^2}{24}[-3G'(\omega) + 3G'(2\omega) - G'(3\omega)] \qquad (26)$$

$$[v_3](\omega) = \frac{G''_{33}}{\omega} = \frac{1-a^2}{8} \int_0^\infty G(s)[\cos\omega s - 2\cos 2\omega s + \cos 3\omega s]ds$$
$$= \frac{1-a^2}{24\omega}[3G''(\omega) - 3G''(2\omega) + G''(3\omega)]. \qquad (27)$$

Equations (24)–(27) represent the first of three major theoretical contributions of our work here. As mentioned in the introduction, recent results of (Song et al. 2020) (Eq.(6) in that work) can be retrieved as a subset of our Equations (24)–(27) by assuming a single mode Maxwell effective relaxation modulus (also note a difference in notation, where their $\zeta$ is related to our $a$ as $\zeta = 1 - a$).

Another material function defined to study the MAOS regime was introduced by Hyun and Wilhelm (Hyun and Wilhelm 2009) and it is related to the Chebyshev metrics above (Hyun et al. 2011; Ewoldt and Bharadwaj 2013) by

$$Q_0 = \frac{\sqrt{[e_3]^2 + (\omega[v_3])^2}}{\sqrt{G'^2 + G''^2}}. \qquad (28)$$

The results of this work can be used to calculate this MAOS measure $Q_0$, but we prefer using Eqs. (24)–(27) as the MOAS material functions to study the independent contributions of the different nonlinearities present in the system, which have distinct interpretations (Ewoldt and Bharadwaj 2013).

It follows from the equations (24)–(27) that the MAOS material functions are linear combinations of the linear viscoelastic storage and loss moduli, evaluated at different frequencies. This is a general feature of time-strain separable MAOS signatures (Martinetti and Ewoldt 2019), as will be discussed in Section I.A. Furthermore, the above expressions allow us to compute oscillatory shear material functions as a function of the effective time-dependent relaxation modulus $G(s)$ and the affinity parameter $a$. In addition, the term $1-a^2$ appears as a front factor and changes the magnitudes of the nonlinearities, but not their signs. Therefore, as expected, the JS/GS model is only able to predict shear thinning, not thickening.

A physically meaningful range for $a$ is $0 < a \leq 1$, and considering the extremes of this limit is insightful. Although mathematically $a$ can take any value, it is unreasonable to assume that the stress causing rate of strain tensor $\underline{\underline{\dot{\gamma}}}_{NA}(t') = a\underline{\underline{\dot{\gamma}}}(t')$ is greater than the imposed rate or of opposite sign to it. Furthermore, the visualization of streamlines and material deformation, based on results of Section IV.B, for $a < 0$ generate physically unreasonable deformation (see Appendix A). The first extreme of the physical range, $a = 1$, which corresponds to the affine limiting case, is incapable of predicting shear stress nonlinearities, confirmed with all four nonlinearities vanishing at $1 - a^2 = 0$. This result agrees with the fact that in the affine limit, the JS/GS model reduces to the UCM model, which is known to have no shear thinning. On the other hand, the limit of $a \to 0$ corresponds to the Corotational Maxwell model (Giacomin et al. 2011) of Goddard and Miller and has the most nonlinear behavior, with a deformation history consisting of pure rotation, as would be shown in section B. As Johnson-Segalman (Johnson and Segalman 1977) note in their work, $aG_A(s)$ should stay non-zero as that limit is taken, otherwise the stress tensor will go to zero as well.

## B. Single-mode response

From the general MAOS solution for any relaxation function, we now explore single mode exponential relaxation to illustrate the behavior of the model, with

$$G(s) = G_0 \exp(-s/\tau_0), \tag{29}$$

where $G_0$ is the elastic modulus magnitude and $\tau_0$ is the relaxation time. While $G(s)$ can take many different mathematical forms, the single-mode relaxation is a building block to understand more complex relaxation spectra, as $G(s)$ can typically be approximated by a sum of exponential modes (Prony series).

The MAOS material functions for the single-mode JS/GS response are found by substituting Eq.(29) into Eqs.(24)–(27),

$$G'(\omega) = G_0 \frac{\text{De}^2}{1+\text{De}^2}$$
$$G''(\omega) = G_0 \frac{\text{De}}{1+\text{De}^2}$$
$$[e_1](\omega) = \frac{a^2-1}{6} G_0 \frac{9\text{De}^4}{(1+\text{De}^2)(1+4\text{De}^2)}$$
$$[v_1](\omega) = \frac{a^2-1}{12} G_0 \tau_0 \frac{9\text{De}^2}{(1+\text{De}^2)(1+4\text{De}^2)} \quad (30)$$
$$[e_3](\omega) = \frac{a^2-1}{6} G_0 \frac{9\text{De}^4(\text{De}^2-1)}{(1+\text{De}^2)(1+4\text{De}^2)(1+9\text{De}^2)}$$
$$[v_3](\omega) = \frac{a^2-1}{12} G_0 \tau_0 \frac{3\text{De}^2(1-11\text{De}^2)}{(1+\text{De}^2)(1+4\text{De}^2)(1+9\text{De}^2)}$$

where the Deborah number, $\text{De} = \omega \tau_0$, is a dimensionless measure of frequency[§,**]. These fingerprints are shown in Figure 2(a)–(b) as a function of the nonlinear parameter $a$, which can be compared to other MAOS signatures as surveyed in (Bharadwaj and Ewoldt 2015a).

The signs and shapes of the MAOS material functions in Figure 2b hold important physical interpretation (Ewoldt and Bharadwaj 2013). First, it is clear that for any relaxation modulus considered, the model shows elastic softening and viscous thinning based on the negative sign of $[e_1](\omega)$ and $[v_1](\omega)$ across all time-scales. Moreover, the sign change present in $[e_3](\omega)$ means that as the Deborah number is increased above $\text{De} = 1$, the elastic softening will be driven by large instantaneous strain rather than large rate-of-strain. Similarly, for $[v_3](\omega)$, the viscous thinning is driven by large strains for $\text{De} > 0.3$, and by large strain rate for lower Deborah numbers. The frequency-dependence (e.g. location of sign changes) is independent of both $G_0$ and $a$, whereas the magnitudes of the nonlinear MAOS functions are all linearly proportional to $G_0$ and further depend on the affinity parameter $a$ as shown in Figure 2. We re-emphasize here that although $a \to 0$ shows the highest magnitude of nonlinearity, the affine relaxation modulus of the material has to go to infinity in this limit which is a non-physical assumption. Combining modes with different time scales can introduce the possibility of more than one sign change in $[e_3](\omega)$ and $[v_3](\omega)$, which can be observed in continuous spectra models as well. The reader is referred to the work of Martinetti & Ewoldt for the shapes introduced by those relaxation spectra (Martinetti and Ewoldt 2019).

---

[§] (Gordon and Everage 1971) derived analytical expressions for the first harmonics $\sigma_1'$ and $\sigma_1''$ in the LAOS regime for the single mode relaxation limit. These expressions can be expanded to obtain $[e_1]$ and $[v_1]$ in Eq.(30).

[**] We note that a more complicated route to Eq.(30) could have started from the fully nonlinear LAOS analytical results of the Oldroyd 8-constant model, which the single-mode JS/GS is a subset of, derived by (Saengow et al. 2017). In that work, the fully nonlinear (not truncated) oscillatory shear stress solution is given for single mode relaxation. The JS/GS limit of the Oldroyd 8-constant model could be taken, and this stress can then in theory be expanded around $\gamma_0 = 0$ to obtain the single-mode MAOS material functions in Eq.(30).

While the results in terms of effective modulus $G(t)$ are useful for fitting experimental data, it is insightful to consider the *affine* relaxation modulus $G_A(t)$ (see Eq.(23)). In Appendix B, we analyze this perspective and study how the nonlinear MAOS functions change with the affinity parameter $a$ while keeping the affine modulus $G_A(t)$ fixed. Interestingly, although decreasing $a$ increases the nonaffinity and hence introduces more nonlinearity in the material element deformation (Section IV.B), for fixed affine relaxation modulus, the slip first increases but then *decreases* the magnitude of the resulting nonlinear functions. This perspective is important for microstructural interpretation, e.g. if slip is truly the cause of the nonlinearity, then for a fixed affine relaxation modulus, the maximum nonlinearity occurs with the slip parameter $a = \sqrt{1/3}$, and not $a = 0$.

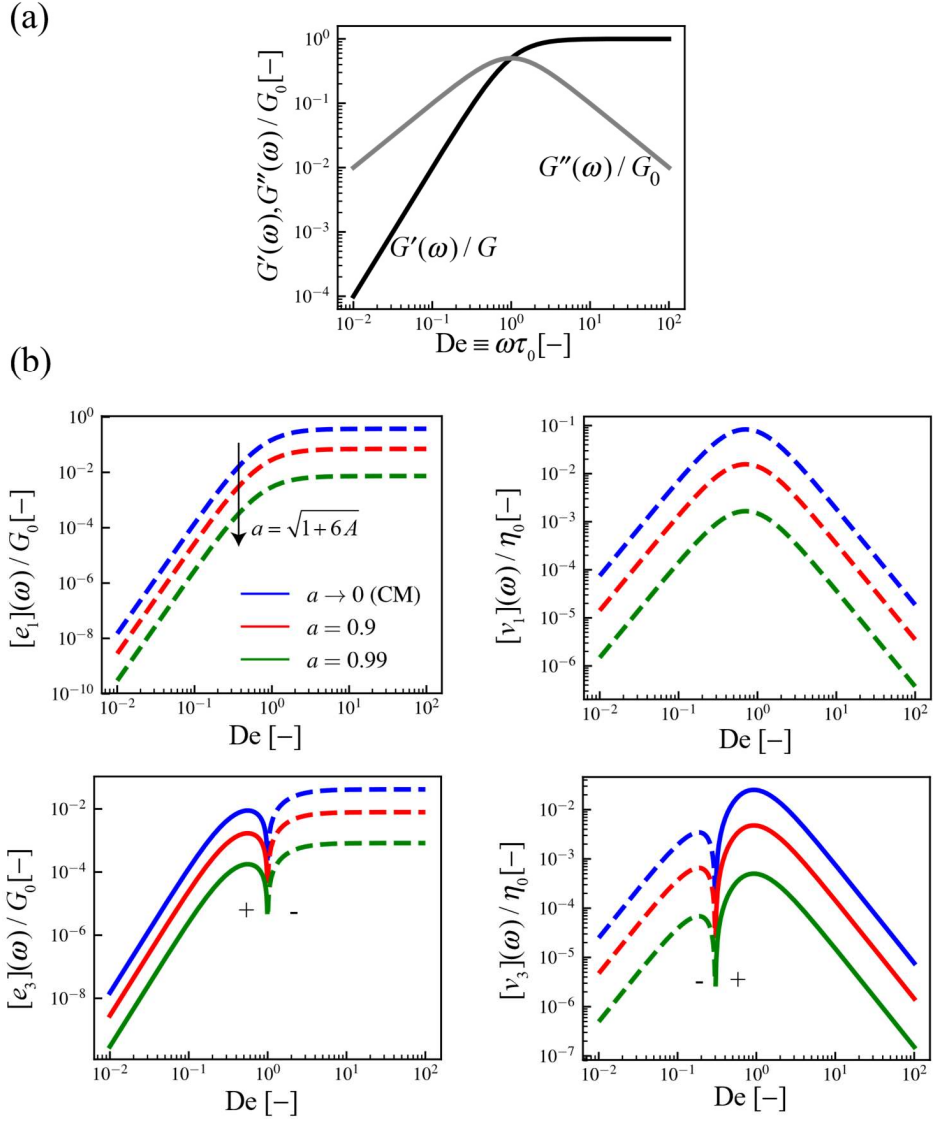

Figure 2: Material functions for (a) SAOS and (b) MAOS of the Johnson-Segalman/Gordon-Schowalter model obtained for single-mode Maxwell relaxation. The elastic intrinsic measures are normalized by the effective modulus $G_0$, and the viscous measures by $\eta_0 = G_0 \tau_0$.

## IV. DISCUSSION

### A. Time-Strain Separability

In recent work, Martinetti and Ewoldt derived the general functional form of the MAOS material functions of a time-strain separable (TSS) model for viscoelastic fluids. A model is time strain separable if the stress can be written as

$$\sigma_{21}(t) = \int_{-\infty}^{t} m(t-t')h(\gamma)\gamma \, dt' \qquad (31)$$

where $m(t) = -\dfrac{dG(t)}{dt}$ and $\gamma = \gamma(t,t')$ is the accumulated strain between time $t'$ to $t$. In the weakly-nonlinear limit, the damping function $h$ can be expanded as

$$h(\gamma) = 1 + A\gamma^2 + O(\gamma^4). \qquad (32)$$

Given this general TSS form, the MAOS functions will be a linear combination of the linear viscoelastic functions (Eqs.27(a)–(d) in (Martinetti and Ewoldt 2019))

$$[e_1](\omega) = -\frac{3}{2}A[G'(\omega) + G'(2\omega)], \qquad (33)$$

$$[v_1](\omega) = -\frac{3}{2}A[G''(\omega) + G''(2\omega)], \qquad (34)$$

$$[e_3](\omega) = -\frac{3}{2}A[G'(\omega) + G'(2\omega) + G'(3\omega)], \qquad (35)$$

$$[v_3](\omega) = -\frac{3}{2}A[G''(\omega) + G''(2\omega) + G''(3\omega)]. \qquad (36)$$

Comparing these Eqs.(33)–(36) to the results of solving the JS/GS model in Eq.(30) proves that the JS/GS model belongs to the TSS MAOS class. Furthermore, the nonlinear parameter $A$ is related to the JS/GS nonlinear parameter as

$$A = \frac{a^2 - 1}{6}. \qquad (37)$$

Equation (37) represents the second of three major theoretical contributions of our work here. This result shows that all the properties of the general TSS MAOS result discussed in (Martinetti and Ewoldt 2019) apply to the JS/GS model solution derived here. More importantly, Eq.(37) provides a physical interpretation (in terms of non-affine deformation) for the TSS parameter $A$, within the range $-\frac{1}{6} < A < 0$ for $0 < a < 1$.

Table 1: Interpretations available for the TSS MAOS nonlinearity parameter $A$, for several constitutive models showing TSS in MAOS. Prior results provide interpretations for only specific values of $A$, whereas the work here covers a range.

| Constitutive model | $A$ | Interpretation |
|---|---|---|
| TSS | $A \in \mathbb{R}$ | N/A |
| JS/GS | $-0.1\overline{6} \leq A < 0$ | Non-affine deformation |
| Corotational Maxwell(CM)(Giacomin et al. 2011) | $-0.1\overline{6}$ | Pure rotation of material elements having infinite relaxation mode strength |
| Doi-Edwards | -0.238 | See (Doi 1980) |
| Doi-Edwards IA[a,] | -0.454 | See (Doi and Edwards 1978) |
| L-MSF[b,] | -0.138 | See (Wagner et al. 2001) |
| Q-MSF[c,] | -0.038 | See (Wagner et al. 2001) |

[a] Doi-Edwards with independent alignment (IA) approximation.
[b] Linear molecular stress function (L-MSF).
[c] Quadratic molecular stress function (Q-MSF).

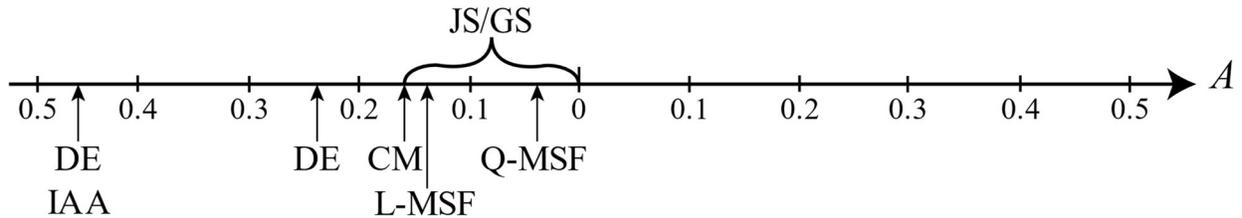

Figure 3: Visualization of Table 1 information on a number-line for context. The JS/GS results here provide an interpretation for a specific range of the nonlinear TSS MAOS parameter $A$. Positive values $A > 0$ correspond to TSS shear-thickening/shear-stiffening material.

Contextualizing the JS/GS model with other known MAOS TSS models provides phenomenological understanding of its nonlinear behavior and demonstrates the added ability to interpret TSS data. Table 1 shows a collection of TSS constitutive models. All these models share the same mathematical structure but differ in the value of the nonlinear parameter $A$ and in their physical interpretation. While molecular models, such as Doi-Edwards, and semi-empirical models, such as JS/GS, have a distinct physical interpretation, phenomenological models such as the general TSS model do not. Moreover, when MAOS experimental data is found to be best fit using a TSS mathematical form with a specific $A$ value, material property inference is only possible when this value corresponds to a specific value predicted from the molecular level theory.

The number line of $A$ shown in Figure 3 illustrates which values of $A$ are covered by an available theory with published MAOS solution. The result of this work covers a range of $A$ values which did not have a physical interpretation before. In addition, comparing the magnitude of nonlinearity $A$ of JS/GS with other TSS models shows that it can only be as nonlinear as the corotational Maxwell model. Furthermore, shear stiffening materials have positive values of $A$, e.g. PVA-Borax transient networks can show a range from $A = 0.08$–$0.2$ (from re-interpretation of data in (Martinetti et al. 2018) [††], providing a physical intuition to what the JS/GS model cannot predict. In section V, a case study will be used to illustrate the utility of having an interpretation in this range by using published MAOS data for a linear polymer melt.

Finally, two clarifying notes regarding TSS models in the literature. First, although (Song et al. 2020) added the Larson model and White-Metzner with Carreau viscosity to their list of TSS models, these models are *not* TSS based on the definition of (Martinetti and Ewoldt 2019). This confusion often occurs when a nonlinear parameter appears as a factorized front factor in their MAOS material functions. However, all TSS MAOS functions must have the same *frequency* dependent shapes. Based on the analytic solutions derived in (Song et al. 2020), the MAOS functions do *not* satisfy those mathematical forms (Eqs.(33) – (36)) and therefore are not TSS. The second note is regarding a common fitting approach, where certain relaxation modes used to fit

---

[††] In [(Martinetti et al. 2018)], experimental $[e_3]$ and $[v_3]$ material functions were provided for an extensive range of PVA-Borax compositions. We used here the plateaus of $[e_3]$ at high frequency, reported in their Table 3, to calculate the apparent nonlinear TSS parameter as $A = \frac{2}{9} \lim_{\omega \to \infty} \frac{[e_3](\omega)}{G_0}$ using Eqn. (35) and assuming a high frequency plateau $G_0$.

SAOS data are neglected when computing MAOS nonlinear signals, as if the nonlinear parameter is not constant, but depends on the associated relaxation timescale. In this case, even if the model in its single mode form is TSS, the considered model fit is not TSS, since Eqs (33) – (36) will not hold anymore.

## B. Visualizing Non-affine Deformation

Understanding the evolution of a material element in the non-affine deformation model of JS/GS is critical to connect it to a clear molecular picture. Therefore, the goal of this section is to visualize this evolution by analyzing the model equations presented in Sections II and III, and this forms the third major contribution of this work. This visualization is relevant to models outside the JS/GS family, such the Phan-Thien/Tanner (PPT) (Thien and Tanner 1977), where the same non-affine deformation is assumed but with a different stress calculator. We note that previous attempts have been made to visualize the deformation history of the JS/GS model by (Petrie 1979), where the trajectory of a point is tracked back in time. Here, we compute the effective non-affine velocity field $\underline{v}_{NA}(\underline{x},t)$ and show its Eulerian streamlines and Lagrangian pathlines, which allow us to track the deformation of material elements in the non-affine model forward and backward in time.

The non-affine slipping motion of the JS/GS model can be visualized by considering the effective velocity gradient $\underline{\underline{L}}$ in Eq.(6). We write this as

$$\underline{\underline{L}} = \underline{\nabla} \underline{v}_{NA}(\underline{x},t) \tag{38}$$

where we interpret $\underline{v}_{NA}(\underline{x},t)$ as the non-affine velocity field. Then, for an assumed continuum flow field $\underline{v}(\underline{x},t)$, Eq.(6) is used to compute $\underline{\underline{L}}$ and Eq.(38) is integrated to find $\underline{v}_{NA}(\underline{x},t)$. Here we show two visualizations of $\underline{v}_{NA}(\underline{x},t)$: Eulerian streamlines and Lagrangian material element cubes.

Assuming the homogeneous simple shear flow defined in Eq. (7), it follows from Eq.(6) that the velocity gradient is given by

$$\underline{\underline{L}} = \underline{\nabla} \underline{v}_{NA}(\underline{x},t) = \begin{pmatrix} 0 & \frac{a+1}{2}\dot{\gamma}(t) & 0 \\ \frac{a-1}{2}\dot{\gamma}(t) & 0 & 0 \\ 0 & 0 & 0 \end{pmatrix} \quad (39)$$

where $\dot{\gamma}(t)$ can have any time dependence. As expected for homogeneous simple shear flow, the velocity gradient is independent of position $\underline{x}$, and this is maintained in the JS/GS model. Let $\underline{v}_{NA}(\underline{x},t)$ be one possible velocity field giving rise to stress in the JS/GS model satisfying Eq.(39). It follows that the $x$-component of the velocity field $v_{NA,x}$ is defined by three differential equations which are $\frac{\partial v_{NA,x}}{\partial y} = \frac{a+1}{2}\dot{\gamma}(t)$ and $\frac{\partial v_{NA,x}}{\partial x} = \frac{\partial v_{NA,x}}{\partial z} = 0$. Similarly the $y$-component is defined by $\frac{\partial v_{NA,y}}{\partial x} = \frac{a-1}{2}\dot{\gamma}(t)$ and $\frac{\partial v_{NA,y}}{\partial y} = \frac{\partial v_{NA,y}}{\partial z} = 0$ and the $z$-component by $\frac{\partial v_{NA,z}}{\partial x} = \frac{\partial v_{NA,z}}{\partial y} = \frac{\partial v_{NA,z}}{\partial z} = 0$. The general form of the velocity field satisfying these equations is

$$\underline{v}_{NA}(\underline{x},t) = \begin{bmatrix} \frac{a+1}{2}\dot{\gamma}(t)y + C_1(t) \\ \frac{a-1}{2}\dot{\gamma}(t)x + C_2(t) \\ C_3(t) \end{bmatrix}. \quad (40)$$

To observe the deformation of a material element without any translation we choose a fixed zero velocity at the origin, $\underline{v}_{NA}(\underline{0},t) = 0$, to obtain

$$\underline{v}_{NA}(\underline{x},t) = \begin{bmatrix} \frac{a+1}{2}\dot{\gamma}(t)y \\ \frac{a-1}{2}\dot{\gamma}(t)x \\ 0 \end{bmatrix}. \quad (41)$$

The Eulerian streamlines are defined by

$$\frac{dy}{dx} = \frac{v_{\text{NA},y}}{v_{\text{NA},x}} = \frac{a-1}{a+1}\frac{x}{y}. \quad (42)$$

The equation of the streamline passing through any point on the $y$-axis, $(0, y_{\text{int}})$, is given by

$$\frac{1-a}{1+a}\frac{x^2}{y_{\text{int}}^2} + \frac{y^2}{y_{\text{int}}^2} = 1 \quad (43)$$

where $y_{\text{int}} \in \mathbb{R}$. The shapes of the streamlines are self-similar and independent of the strain or strain rate for a particular value $y_{\text{int}}$ for this flow field. Nevertheless, the value of the stream function across each of the streamlines will depend on the instantaneous strain rate.

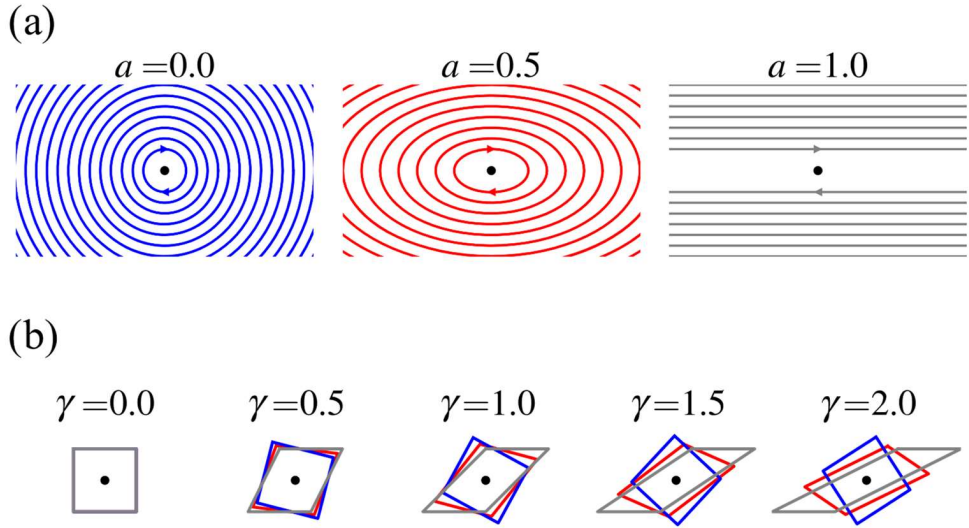

Figure 4: Non-affine flow visualization in simple shear. (a) Eulerian streamlines, and (b) Lagrangian material element deformation, for the $a = 0.0$ corotational (blue), $a = 0.5$ (red), and $a = 1.0$ affine (gray). The streamlines are plotted for $y_{\text{int}} = \{1.0, 1.5, 2.0, ...\}$. Since the corotational limit is purely rotational, the unit square is not deformed and can be used for visual reference

Figure 4a shows derived streamlines for the case of $a = 0.5$ and the two limiting cases of the model. For all values of $0 < a < 1$, the streamlines will be ellipses. However, as $a$ approaches the limit of affine motion, $a = 1$, the streamlines approach straight lines, as would be expected for affine simple shear flow. On the other hand, as $a \rightarrow 0$, the streamlines become circles, which is the behavior of the corotational model. As mentioned in Section A, the negative *a* values are physically unrealistic, which is further proven by the associated non-affine streamlines and deformation shown in Appendix A.

A second and perhaps more useful visualization is of Lagrangian material elements deformed by the field $\underline{v}_{NA}(\underline{x},t)$, as shown in Figure 4b. For this, consider cube-shaped elements centered at the origin, which omits any translation and isolates the deformation of the elements. The path of each material point $\underline{x}_i(t)$ is defined by the differential equation

$$\frac{d\underline{x}_i(t)}{dt} = \underline{v}_{NA}(\underline{x}_i(t),t) \tag{44}$$

which can be integrated (numerically if needed) starting from any specific position $\underline{x}_i(0)$. In our case, starting from multiple positions that define the boundaries of an initial material volume. The initial size of the material cube does not affect the relative shape of the deformation, due to the self-similar streamlines and pathlines.

For homogeneous simple shear, $\underline{v}_{NA}(\underline{x},t)$ from Eq.(41) results in zero velocity in the *z*-direction, therefore a two-dimensional projection of the cube is sufficient. Starting from an undeformed unit square in the *x-y* plane, we integrate material points along the pathlines based on Eq.(44), taking snapshots at different points in time, which map to different values of imposed macroscopic shear strain $\gamma(t) = \int_0^t \dot{\gamma}(t')dt'$. The time history of the shear rate does not affect the result, because the streamlines are independent of time (Eqn.(43)) for this flow field, which makes the deformation dependent only on the accumulated strain.

Figure 4b shows how the element changes as the strain increases for three cases: affine (*a* = 1), corotational (*a* = 0), and *a* = 0.5. In the corotational limit, material stress elements are only rotated, keeping the unit square shape intact. On other hand, in the affine limit, the square deforms along the horizontal streamlines defined by the affine velocity field. In the interesting range introduced by the JS/GS model, the material elements are stretched and rotated, depending on the affinity parameter *a*. Figure 4b shows that for small strains, the deviation of the non-affine history from the affine limit is small and grows gradually as the strain amplitude is increased. Similarly, the deviation of the non-affine stress prediction compared to the linear affine limit increases as the amplitude is increased as illustrated in Figure 1 due to the deviation of the material element deformation.

The results presented here can be generalized for other types of flow fields in extension, shear, or combination. First, the same methodology can be used to solve for the non-affine velocity field and material element deformation in other flow fields. For example, with uniaxial extension the non-affine velocity field is $\underline{v}_{NA} = a\underline{v}$. This pure slipping behavior does not qualitatively change how the material elements deform, but it makes it slower. In a similar fashion, different input velocity fields can be analyzed. Second, the deformation illustrated above under a globally homogenous velocity gradient is equivalent to the local deformation of infinitesimal material elements under a nonhomogeneous flow field. This result shows us the general behavior of material elements for a specific velocity gradient for this non-affine deformation model. In the next subsection, the implication of this behavior on the interpretability of the model is discussed.

## C. Model Interpratation and Applicability to Different Material Classes

As mentioned throughout this work, the importance of the JS/GS model is that it offers a possibility for material-level inference from MAOS data. Nevertheless, the material elements, whose non-affine deformation was demonstrated above in Figure 4b, do not correspond to a specific physical or molecular picture. Hence, the JS/GS model is a semiempirical model that consists of two components – the Lodge-like linear viscoelastic stress calculator determined by the affine relaxation modulus $G_A(t)$ and the non-affine material element deformation determined by the affinity parameter $a$. Moreover, it is the nonaffinity that generates the nonlinearity in this model, and it corresponds to a specific physical picture. On the other hand, the affine relaxation modulus can be related to molecular parameters through any known molecular model that satisfies the Lodge-like mathematical form. Examples include but are not restricted to the Green and Tobolsky model of transient polymer networks and the Rouse theory for dilute polymer solutions. Moreover, (Winter and Mours 1997) used the Lodge equation with power law relaxation to model the linear viscoelastic response of critical gels. Consider that the JS/GS model can be used to fit MAOS data of a polymer solution with a specific $G(s)$ and $a$. In this case, the affine relaxation function $G_A(s) = G(s)/a$ can be used to infer the quantitative properties of the polymer chains and solvent in the Rouse model. However, the important addition given by JS/GS is that *a* can be used to infer how the material elements are deforming (Section IV.b).

The utility of the JS/GS model is not restricted to one microstructure or material. Therefore, if a new material satisfies Lodge-like behavior described above, it might be modeled using the JS/GS model, to have a non-affine deformation interpretation. However, the JS/GS model cannot be used for materials that exhibit nonlinear behavior in a form other than non-affine deformation such as finite extensibility, plasticity, or structure/network breaking.

## V.    CASE STUDY

Here we demonstrate the applicability of the JS/GS MAOS predictions to experimental MAOS data already in the public domain. Although several such studies are available, here we use published data for a linear polymer melt as described in the work of (Singh 2019). The fitting results there are adopted as is but can be given a new interpretation based on the JS/GS model, which has the highest credibility score[‡‡] of all the models considered in the original work.

The analysis done in (Singh 2019) shows that between the extensive list of TSS and non-TSS models considered, the most credible fit was found to be a TSS model having a fractional Maxwell spectrum and a nonlinear parameter $A$ = -0.115. This value was interpreted by considering values of available constitutive models (Table 1, Figure 3), but no exact match was found, since the JS/GS results here and in (Song et al. 2020) were not yet available.

---

[‡‡] (Singh 2019) used an effective Bayesian Information Criterion (BIC) $\mathrm{BIC} = RSS_{\min} + N_p \log(N_d)$ to evaluate model credibility by penalizing excessive parameters, and approximation of the full Bayes factors(Freund and Ewoldt 2015), where $RSS_{\min}$ is the root sum of squares of error between the model and the data at the $N_d$ data points for a model having $N_p$ number of parameters.

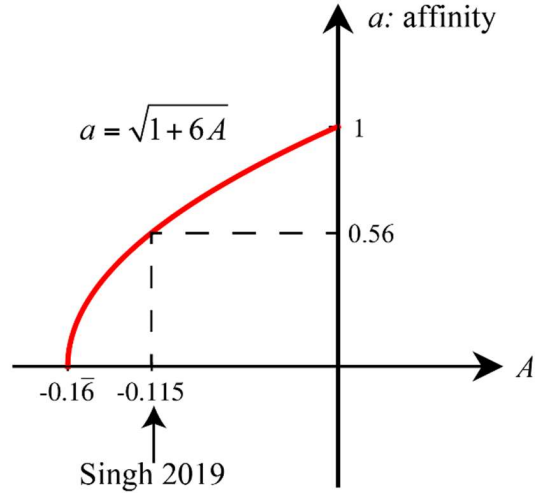

Figure 5: The affinity parameter $a$ as a function of TSS nonlinear parameter $A$. The result of (Singh 2019), corresponding to the best fit of MAOS data taken for a linear polymer melt and falling in the JS/GS range, is shown on the plot.

Now, using the key result of Eq.(37), that value of $A$ can be interpreted as non-affine slip in the context of the JS/GS model. Figure 5 shows the relation between the TSS nonlinear parameter $A$ and the JS/GS affinity parameter $a$, based on Eq.(37). It further locates the result of $A = -0.115$, which corresponds to a JS/GS model having an affinity parameter $a = 0.56$. Moreover, the referenced work also shows fitting results for different $G(s)$ forms, and all of them had an optimal BIC for a value of $A$ that falls in the JS/GS range. In short, this is a clear illustration of the ability of the JS/GS model to fit MAOS data and assign to it a physical interpretation.

## VI. CONCLUSION

The MAOS regime viscoelastic response of the Johnson-Segalman/Gordon-Schowalter non-affine deformation model has been derived for a general relaxation function. The model offers a physical interpretation of a subset of the time-strain separable class which was not previously identified. The molecular picture of non-affine motion was illustrated and visualized in terms of an effective non-affine flow field giving rise to stress. This result is related to previously derived MAOS solutions of the corotational Maxwell model, which is one limit of the JS/GS model. In particular, the molecular picture of the corotational Maxwell model from this perspective can be understood to be that of pure rotation of material elements that have an infinite relaxation modulus. Although the single mode analytical results were derived previously, the thorough study of the generalized integral model and its non-affine deformation presented here contribute to utilizing it in material inference.

One known limitation of the JS/GS model is that it fails for large strains as detailed by Petrie and Larson (Petrie 1979; Larson 1988), and hence it is expected that the model will fail in the LAOS regime. Nevertheless, this limitation does not affect the utility of the model in explaining data in the MAOS regime, since MAOS is a power expansion defined in the limit of strain going to zero. In fact, this work makes it possible to reinterpret experimental data. The case study of Section V demonstrates how the JS/GS model may offer the most credible fit/explanation among many other considered models for MAOS data, suggesting the importance of non-affine deformation in the initial growth of nonlinearities. Nevertheless, credibility of MAOS inference can be further strengthened as more theories and explanations are brought to the realm of MAOS measurements.

The JS/GS picture of non-affine motion is applicable to many material classes (Larson 1988), and holds promise in explaining experimental data. Experimental studies show the presence of non-affine deformation in materials such as polymer hydrogels, synthetic polymers, and other types of soft matter (Wen et al. 2012). Further experiments on these materials can show if the JS/GS can be used to explain the observations, or whether other types of non-affine deformation have to be considered (Rubinstein and Panyukov 1997; Kroon 2011). Comparing our analytical solutions to weakly-nonlinear MAOS experiments should assist in building our understanding of material behavior and to develop more credible non-affine deformation models and material-level inference.

## Acknowledgements

Research supported by the U.S. Department of Energy, Office of Basic Energy Sciences, Division of Materials Sciences and Engineering under Award # DE-SC0020858, through the Materials Research Laboratory at the University of Illinois at Urbana-Champaign.

## APPENDIX A

The unphysical range of affinity parameter $a<0$ produces peculiar streamlines and material element deformation, as we show here. It is known that for $a=-1$ the GS convected derivative is equivalent to the lower-convected derivative and therefore studying this limit is instructive. To do this we follow the same steps taken in Section IV.B, but rewriting the streamline equation, Eq.(43), to show the equation of the streamline passing through a particular x-intercept $(x_{int},0)$, as

$$\frac{x^2}{x_{int}^2} + \frac{1+a}{1-a}\frac{y^2}{x_{int}^2} = 1. \qquad (45)$$

Figure 6 shows the streamlines and deformation for negative affinity parameter and compares it to the case of $a = 1$. For an imposed homogeneous simple shear deformation with non-zero x-velocity (the same as in Figure 4), for $a = -1$ the material is deformed perpendicularly with a non-zero y-velocity. For $a = -0.5$ the material is rotated and deformed, but with the deformation also occurring perpendicularly to the imposed gradient. Both of these cases display unphysical behavior, adding to the reasons of why this limit of the model should not be considered to fit experimental data.

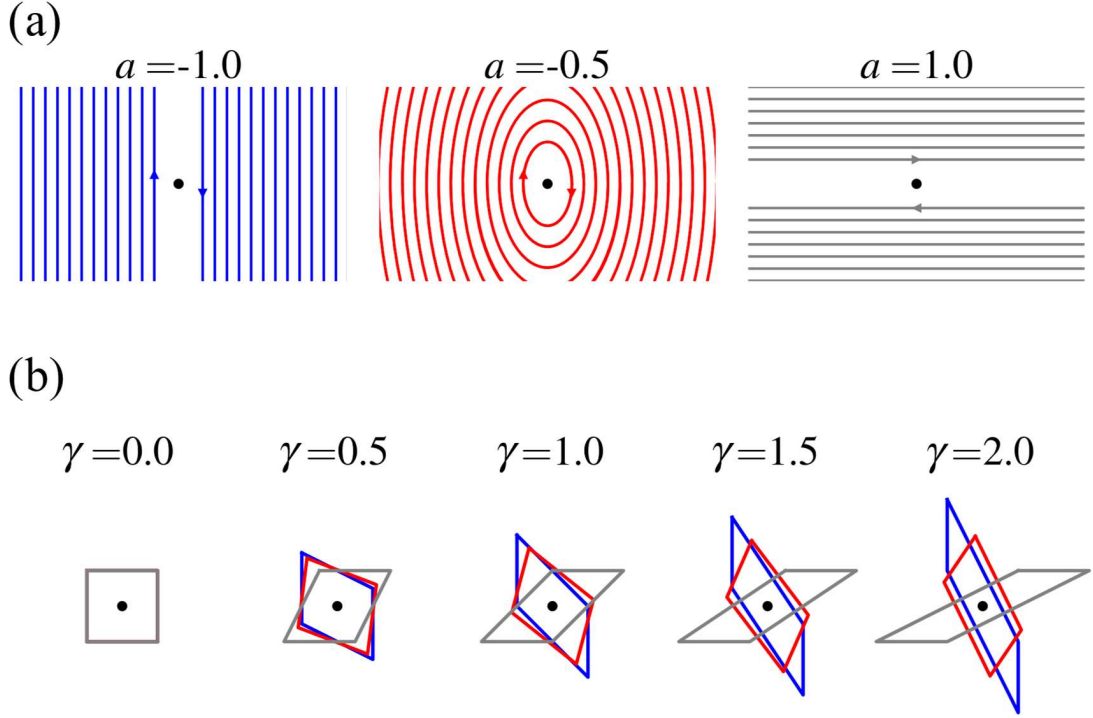

Figure 6: Non-affine flow visualization in simple shear. (a) Eulerian streamlines, and (b) Lagrangian material element deformation, for the $a = -0.5$ (red), $a = -1$ (blue), and $a = 1.0$ affine (gray). The streamlines are plotted for $x_{int} = \{1.0, 1.5, 2.0, ...\}$ when $a = \{-1, -0.5\}$ and $y_{int} = \{1.0, 1.5, 2.0, ...\}$ for $a=1$.

## APPENDIX B

Peculiar results occur if we conceptualize a fixed *affine* modulus $G_A(s)$ while changing the slip parameter $a$, as if the underlying material structure is the same and we only change the amount of slip. In this case the MAOS nonlinear strength is a non-monotonic function of $a$, in contrast to the monotonic trend with $a$ as shown in Figure 2 for a fixed *effective* modulus $G(s)$. Fixing the effective relaxation modulus $G(s) = aG_A(s)$, as we analyzed in Section III.B, is perhaps more relevant for experiments since it is measured directly. Nevertheless, fixing $G(s)$ hides the non-affine conceptual picture of the microstructural nonlinearities, and we will see how fixing $G_A(s)$ changes the analysis.

Starting from the general MAOS solution for any relaxation function (Section III.A), we consider a single mode exponential relaxation for the affine modulus

$$G_A(s) = G_{A,0} \exp(-s/\tau_0), \tag{46}$$

where $G_{A,0}$ is the affine elastic modulus magnitude and $\tau_0$ is the relaxation time. The resulting MAOS material functions are

$$\begin{aligned}
G'(\omega) &= (aG_{A,0}) \frac{\text{De}^2}{1+\text{De}^2} \\
G''(\omega) &= (aG_{A,0}) \frac{\text{De}}{1+\text{De}^2} \\
[e_1](\omega) &= \frac{a^2-1}{6}(aG_{A,0}) \frac{9\text{De}^4}{(1+\text{De}^2)(1+4\text{De}^2)} \\
[v_1](\omega) &= \frac{a^2-1}{12}(aG_{A,0})\tau_0 \frac{9\text{De}^2}{(1+\text{De}^2)(1+4\text{De}^2)} \\
[e_3](\omega) &= \frac{a^2-1}{6}(aG_{A,0}) \frac{9\text{De}^4(\text{De}^2-1)}{(1+\text{De}^2)(1+4\text{De}^2)(1+9\text{De}^2)} \\
[v_3](\omega) &= \frac{a^2-1}{12}(aG_{A,0})\tau_0 \frac{3\text{De}^2(1-11\text{De}^2)}{(1+\text{De}^2)(1+4\text{De}^2)(1+9\text{De}^2)}
\end{aligned} \tag{47}$$

The results from Eq. (47) are plotted in Figures 7a-b. Two differences with Figure 2 stand out. First, the linear moduli have a linear dependence on the affinity parameter, where both moduli are maximum for the affine limit ($a=1$) and vanish in the corotational limit ($a=0$). Second, the dependence of the magnitude of the nonlinear functions has a non-monotonic dependence on the affinity parameter, unlike what was observed in Figure 2b. By examining Eq.(47), the dependence of the nonlinearity strength on $a$ can be captured by the factor $\left|\frac{a(a^2-1)}{6}\right|$, compared to $\left|\frac{a^2-1}{6}\right|$. for the fixed $G(s)$ case from Eq. (30). This non-monotonic dependence is illustrated in Figure 7c. Although the slip increases the nonlinearity of the deformation response, the stress response is more complicated. The slip introduced by non-affinity decreases the effective linear modulus of the material and therefore weakens the overall stress response. Thus, the maximum magnitude of the nonlinear functions for this model will be observed for $a = \sqrt{\frac{1}{3}} \approx 0.577$ for a fixed affine relaxation function. Using a fixed effective modulus hides this finding.

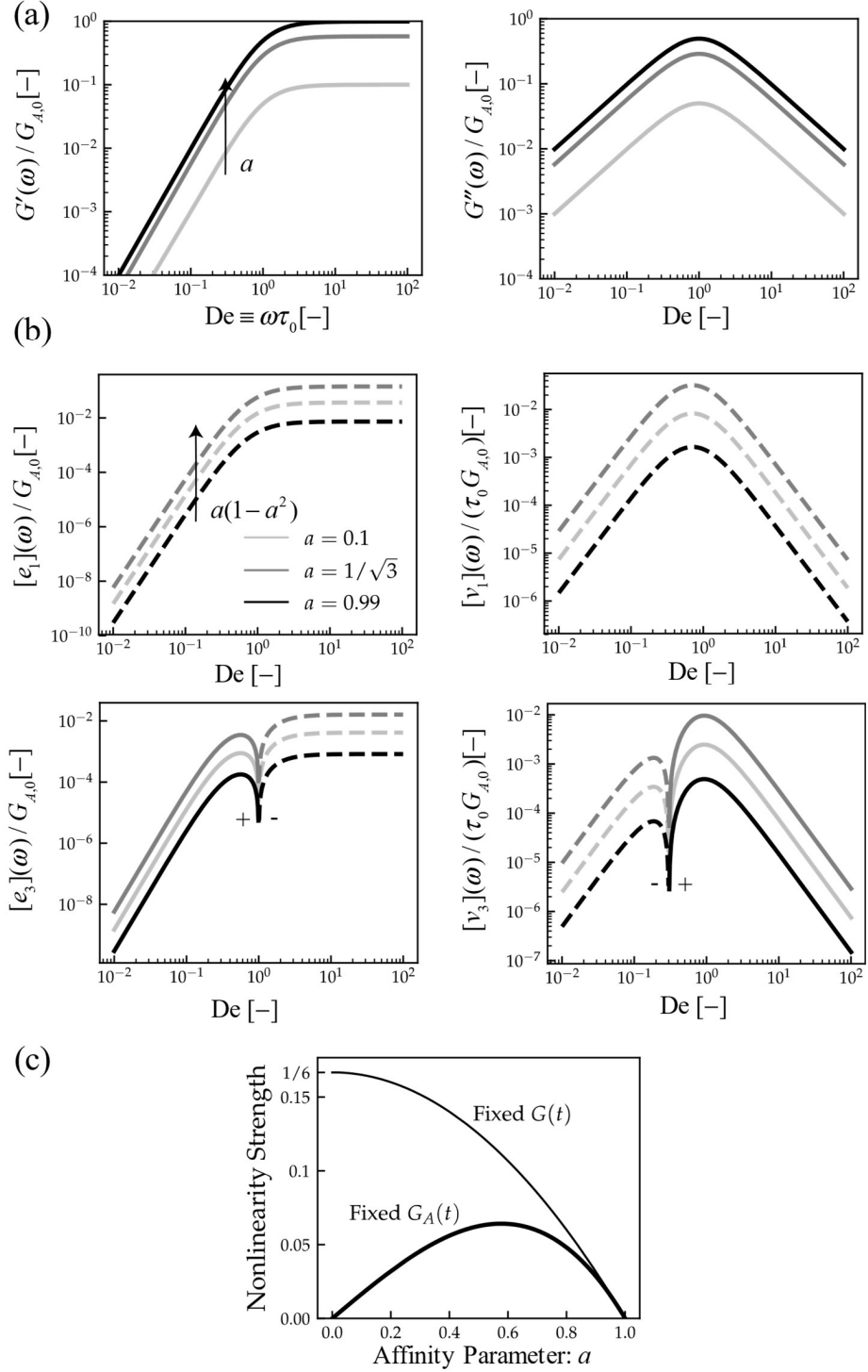

Figure 7: Material functions normalized by the affine modulus strength $G_{A,0}$ for (a) SAOS and (b) MAOS of the Johnson-Segalman/Gordon-Schowalter model with single-mode Maxwell relaxation. Plot (c) indicates how the strength of the MAOS nonlienarities have a non-monotonic dependence on $a$ if the *affine* modulus strength is held fixed, with maximum nonlinear shear stress at $a = \sqrt{\frac{1}{3}} \approx 0.577$.

# References

Bharadwaj NA, Ewoldt RH (2015a) Constitutive model fingerprints in medium-amplitude oscillatory shear. J Rheol 59:557–592. https://doi.org/10.1122/1.4903346

Bharadwaj NA, Ewoldt RH (2015b) Single-point parallel disk correction for asymptotically nonlinear oscillatory shear. Rheol Acta 54:223–233. https://doi.org/10.1007/s00397-014-0824-9

Bharadwaj NA, Ewoldt RH (2014) The general low-frequency prediction for asymptotically nonlinear material functions in oscillatory shear. J Rheol 58:891–910. https://doi.org/10.1122/1.4874344

Bharadwaj NA, Schweizer KS, Ewoldt RH (2017) A strain stiffening theory for transient polymer networks under asymptotically nonlinear oscillatory shear. J Rheol 61:643–665. https://doi.org/10.1122/1.4979368

Blackwell BC, Ewoldt RH (2016) Non-integer asymptotic scaling of a thixotropic-viscoelastic model in large-amplitude oscillatory shear. J Nonnewton Fluid Mech 227:80–89. https://doi.org/10.1016/j.jnnfm.2015.11.009

Chambon F, Winter HH (1987) Linear viscoelasticity at the gel point of a crosslinking PDMS with imbalanced stoichiometry. J Rheol 31:683–697. https://doi.org/10.1122/1.549955

Cho KS, Hyun K, Ahn KH, Lee SJ (2005) A geometrical interpretation of large amplitude oscillatory shear response. J Rheol 49:747–758. https://doi.org/10.1122/1.1895801

Choi J, Nettesheim F, Rogers SA (2019) The unification of disparate rheological measures in oscillatory shearing. Phys Fluids 31:073107. https://doi.org/10.1063/1.5106378

Davis WM, Macosko CW (1978) Nonlinear dynamic mechanical moduli for polycarbonate and PMMA. J Rheol 22:53–71. https://doi.org/10.1122/1.549500

Doi M (1980) A constitutive equation derived from the model of doi and edwards for concentrated polymer solutions and polymer melts. J Polym Sci Polym Phys Ed 18:2055–2067. https://doi.org/10.1002/pol.1980.180181005

Doi M, Edwards SF (1978) Dynamics of concentrated polymer systems. Part 1.—Brownian motion in the equilibrium state. J Chem Soc, Faraday Trans 2 74:1789–1801.

https://doi.org/10.1039/F29787401789

Ewoldt RH (2013) Defining nonlinear rheological material functions for oscillatory shear. J Rheol 57:177–195. https://doi.org/10.1122/1.4764498

Ewoldt RH, Bharadwaj NA (2013) Low-dimensional intrinsic material functions for nonlinear viscoelasticity. Rheol Acta 52:201–219. https://doi.org/10.1007/s00397-013-0686-6

Ewoldt RH, Hosoi AE, McKinley GH (2008) New measures for characterizing nonlinear viscoelasticity in large amplitude oscillatory shear. J Rheol 52:1427–1458. https://doi.org/10.1122/1.2970095

Freund JB, Ewoldt RH (2015) Quantitative rheological model selection: Good fits versus credible models using Bayesian inference. J Rheol 59:667–701. https://doi.org/10.1122/1.4915299

Giacomin AJ, Bird RB, Johnson LM, Mix AW (2011) Large-amplitude oscillatory shear flow from the corotational Maxwell model. J Nonnewton Fluid Mech 166:1081–1099. https://doi.org/10.1016/j.jnnfm.2011.04.002

Gordon RJ, Everage AE (1971) Bead-spring model of dilute polymer solutions: Continuum modifications and an explicit constitutive equation. J Appl Polym Sci 15:1903–1909. https://doi.org/10.1002/app.1971.070150809

Gordon RJ, Schowalter WR (1972) Anisotropic fluid theory: A different approach to the dumbbell theory of dilute polymer solutions. Trans Soc Rheol 16:79–97. https://doi.org/10.1122/1.549256

Hyun K, Wilhelm M (2009) Establishing a new mechanical nonlinear coefficient Q from FT-rheology: First investigation of entangled linear and comb polymer model systems. Macromolecules 42:411–422. https://doi.org/10.1021/ma8017266

Hyun K, Wilhelm M, Klein CO, et al (2011) A review of nonlinear oscillatory shear tests: Analysis and application of large amplitude oscillatory shear (LAOS). Prog Polym Sci 36:1697–1753. https://doi.org/10.1016/j.progpolymsci.2011.02.002

Johnson MW, Segalman D (1977) A model for viscoelastic fluid behavior which allows non-affine deformation. J Nonnewton Fluid Mech 2:255–270. https://doi.org/10.1016/0377-0257(77)80003-7

Kroon M (2011) An 8-chain model for rubber-like materials accounting for non-affine chain deformations and topological constraints. J Elast 102:99–116. https://doi.org/10.1007/s10659-010-9264-7

Larson RG (1988) Constitutive equations for polymer melts and solutions. Elsevier, Stoneham

Lennon KR, McKinley GH, Swan JW (2020) Medium amplitude parallel superposition (MAPS) rheology. Part 1: Mathematical framework and theoretical examples. J Rheol 64:551–579. https://doi.org/10.1122/1.5132693

Martinetti L, Carey-De La Torre O, Schweizer KS, Ewoldt RH (2018) Inferring the Nonlinear Mechanisms of a Reversible Network. Macromolecules 51:8772–8789. https://doi.org/10.1021/acs.macromol.8b01295

Martinetti L, Ewoldt RH (2019) Time-strain separability in medium-amplitude oscillatory shear. Phys Fluids 31:. https://doi.org/10.1063/1.5085025

Natalia I, Ewoldt RH, Koos E (2020) Questioning a fundamental assumption of rheology: Observation of noninteger power expansions. J Rheol 64:625–635. https://doi.org/10.1122/1.5130707

Onogi S, Masuda T, Matsumoto T (1970) Non-linear behavior of viscoelastic materials. I. Disperse systems of polystyrene solution and carbon black. Trans Soc Rheol 14:275–294. https://doi.org/10.1122/1.549190

Paul E (1969) Non-newtonian viscoelastic properties of rodlike molecules in solution: Comment on a paper by kirkwood and plock. J Chem Phys 51:1271–1272. https://doi.org/10.1063/1.1672148

Petrie CJS (1979) Measures of deformation and convected derivatives. J Nonnewton Fluid Mech 5:147–176. https://doi.org/10.1016/0377-0257(79)85010-7

Radulescu O, Olmsted PD (2000) Matched asymptotic solutions for the steady banded flow of the diffusive Johnson–Segalman model in various geometries. J Nonnewton Fluid Mech 91:143–164. https://doi.org/10.1016/S0377-0257(99)00093-2

Rubinstein M, Panyukov S (1997) Nonaffine deformation and elasticity of polymer networks. Macromolecules 30:8036–8044. https://doi.org/10.1021/ma970364k

Saengow C, Giacomin AJ, Kolitawong C (2017) Exact analytical solution for large-amplitude oscillatory shear flow from Oldroyd 8-constant framework: Shear stress. Phys Fluids 29:043101. https://doi.org/10.1063/1.4978959

Schweizer KS, Curro JG (1994) PRISM theory of the structure, thermodynamics, and phase transitions of polymer liquids and alloys. In: Atomistic Modeling of Physical Properties. Springer Berlin Heidelberg, pp 319–377

Singh PK (2019) Rheological inferences with uncertainty quantification. PhD Thesis, University of Illinois at Urbana-Champaign, USA

Song HY, Kong HJ, Kim SY, Hyun K (2020) Evaluating predictability of various constitutive equations for MAOS behavior of entangled polymer solutions. J Rheol 64:673–707. https://doi.org/10.1122/1.5139685

Thien NP, Tanner RI (1977) A new constitutive equation derived from network theory. J Nonnewton Fluid Mech 2:353–365. https://doi.org/10.1016/0377-0257(77)80021-9

Wagner MH, Rolón-Garrido VH, Hyun K, Wilhelm M (2011) Analysis of medium amplitude oscillatory shear data of entangled linear and model comb polymers. J Rheol 55:495–516. https://doi.org/10.1122/1.3553031

Wagner MH, Rubio P, Bastian H (2001) The molecular stress function model for polydisperse polymer melts with dissipative convective constraint release. J Rheol 45:1387–1412. https://doi.org/10.1122/1.1413503

Wen Q, Basu A, Janmey PA, Yodh AG (2012) Non-affine deformations in polymer hydrogels. Soft Matter 8:8039. https://doi.org/10.1039/c2sm25364j

Winter HH, Mours M (1997) Rheology of polymers near liquid-solid transitions. In: Neutron Spin Echo Spectroscopy Viscoelasticity Rheology. Springer Berlin Heidelberg, Berlin, Heidelberg, pp 165–234